# $I = 2$ pion scattering amplitude with Wilson fermions

Rajan Gupta

*T-8, MS-B285, Los Alamos National Laboratory, Los Alamos, NM 87545*

Apoorva Patel

*Supercomputer Education and Research Centre and Centre for Theoretical Studies*
*Indian Institute of Science, Bangalore 560012, India*

Stephen R. Sharpe

*Physics Department, University of Washington, Seattle, WA 98195*

We present an exploratory calculation of the $I = 2$ $\pi - \pi$ scattering amplitude at threshold using Wilson fermions in the quenched approximation, including all the required contractions. We find good agreement with the predictions of chiral perturbation theory even for pions of mass 560-700 MeV. Within the 10% errors, we do not see the onset of the bad chiral behavior expected for Wilson fermions. We also derive rigorous inequalities that apply to 2-particle correlators and as a consequence show that the interaction in the antisymmetric state of two pions has to be attractive.

12 January 1993

## 1. Introduction

In this paper we calculate the $I = 2$ $\pi\pi$ scattering amplitude at threshold using Wilson fermions. The theoretical foundations of the calculation have been established in a series of papers by Lüscher where he shows how the finite volume dependence of the two particle energy levels in a sufficiently large cubic box is related to the scattering amplitude [1] [2]. In the context of lattice field theories, the leading term in an infinite volume expansion was first given in Ref. [3]. The subtleties of the lattice calculations have been described in detail in Ref. [4], where results of a calculation using staggered fermions is presented. We follow closely the notation of this reference.

The calculation of the $I = 2$ $\pi\pi$ scattering amplitude at threshold has a number of simplifying features. In general, scattering amplitudes are complex, and are only related indirectly to the finite volume energy shift. At threshold, however, the amplitude is real, and is directly related to the energy shift. Also, the signal for pions is much better than for other, heavier, mesons, e.g. rhos. Finally, one needs to calculate only quark and gluon exchanges for the $I = 2$ channel, whereas, in general, there are also annihilation diagrams. The latter are both more difficult to calculate numerically, and are affected more strongly by the use of the quenched approximation.

A major motivation for this work is to test the chiral behavior of the scattering amplitude derived long ago by Weinberg [5] using PCAC and current algebra. We use Wilson's formulation of lattice fermions, for which there are lattice artifacts arising from the explicit breaking of chiral symmetry. We can quantify these corrections by comparing lattice results against the PCAC prediction and against results obtained using staggered fermions on the same set of lattices [4].

Our calculation is exploratory in at least three different ways. First, we use the quenched approximation. Second, we only use two different quark masses, neither very light. Thus we can only make rough extrapolations to the chiral limit. And, finally, we have only used one lattice size ($16^3 \times 40$). This means we must assume the finite volume dependence predicted by Refs. [1] and [3], and cannot check the predictions.

In the earlier calculation, Guagnelli, Marinari and Parisi [6] have done a partial calculation of the scattering amplitude using Wilson (and staggered) fermions. They did not include all the contractions which contribute in the $I = 2$ channel, and thus one cannot extract the scattering amplitude from their results.

The article is organized as follows. A brief theoretical overview is given in Section 2 and the methodology and details of the lattices are given in Section 3. The results are presented in Section 4, and in Section 5 we present a model which explains some features of the data. Section 6 gives some conclusions. In the Appendix we derive rigorous inequalities regarding correlators of the type used in this study, in particular we show that the interaction in the anti-symmetric $\pi - \pi$ channel has to be attractive.

## 2. Theoretical background

Lüscher has derived the relationship between pion scattering amplitudes and the energies of two pion state in finite volume [1][2]. The derivation is valid as long as the box (which we take to be cubic) is large enough that its length $L$ exceeds twice the range of interaction. In general, the relationship is complicated, involving scattering amplitudes over a range of energies and in many partial waves. The relationship simplifies, however, if one expands the energies in powers of $1/L$ and keeps only the first few terms. We consider only the lightest two pion state whose energy we denote by $E$. For infinite volume $E = 2m_\pi$, but the energy is shifted by interactions as $L$ is reduced

$$\delta E = E - 2m_\pi = \frac{T}{L^3}\left(1 - c_1\frac{m_\pi T}{4\pi L} + c_2\left(\frac{m_\pi T}{4\pi L}\right)^2\right) + O(L^{-6}) , \qquad (2.1)$$
$$c_1 = -2.837297 , \qquad c_2 = 6.375183 .$$

To the order shown, the energy shift depends only on $T$, which is the (non-relativistically normalized) scattering amplitude at threshold.

We use the non-relativistically normalized amplitude in Eq. (2.1) since this simplifies its physical interpretation, as explained in Ref. [4]. $T$ is related to the relativistically normalized scattering amplitude by $T^R = -(4m_\pi^2)T$, and to the S-wave scattering length by $T = -4\pi a_0/m_\pi$.

We extract $T$ from our numerical data using Eq. (2.1). *A priori* we do not know the size of the $O(L^{-6})$ terms which we are dropping. Our numerical results suggest, however, that the truncation error is small.

Equation (2.1) holds separately for $I = 0$ and 2 two pion states (Bose symmetry forbids $I = 1$ at threshold). We have done the calculation, however, only for the $I = 2$ channel. To understand why, consider the four types of contraction that contribute to a calculation of the two pion energy, shown in Fig. 1. With present computer resources we can calculate only the first two types, which we label the Direct (D) and Crossed (C) diagrams respectively. These are not sufficient to calculate the $I = 0$ amplitude which gets contributions from all four diagrams. For $I = 2$, however, quark-antiquark annihilation is not possible, and the required result is given by the combination $D - C$. Here we are adopting the convention of showing explicitly the minus sign from Fermi statistics, so that the lines in Fig. 1 represent c-number quark propagators.

We can also calculate the combination $D + C$. This does not project onto a definite isospin, but does select a definite representation in a theory with $N_f \geq 4$, $N_f$ being the number of flavors. $D + C$ picks out the $\overline{qq}qq$ representation having no traced indices, and is antisymmetric under the interchange of either quarks or antiquarks. We refer to this representation as the **A** (for $N_f = 4$ it is the **20**), and to the corresponding scattering amplitude as $T(\mathbf{A})$. For arbitrary $N_f > 2$, the combination $D - C$, which we call the **S**, projects onto the representation with no traced indices and is symmetric under quark exchange. The generalization of the $I = 2$ amplitude $T_2$ is thus $T(\mathbf{S})$. In the quenched approximation the amplitudes $T(\mathbf{S})$ and $T(\mathbf{A})$ are independent of $N_f$, because the Wick contractions are always the same. In particular, $T_2 = T(\mathbf{S})$.

An important test of any calculation of pion scattering amplitudes is that they satisfy the constraints of chiral symmetry. In particular, the threshold amplitudes are determined, in the chiral limit, in terms of $f_\pi$ (which is 93 MeV in our normalization) [5][4]

$$\begin{aligned}
4f_\pi^2 T(\mathbf{S}) &= 1 + O(m_\pi^2 \ln m_\pi) \;, \\
4f_\pi^2 T(\mathbf{A}) &= -1 + O(m_\pi^2 \ln m_\pi) \\
&= -4f_\pi^2 T(\mathbf{S}) + O(m_\pi^2 \ln m_\pi) \;.
\end{aligned} \quad (2.2)$$

These results should apply in the quenched approximation, as discussed in Ref. [4]. Since $T(\mathbf{S}) > 0$, two pions in an **S** representation are repelled in the chiral limit, while in the **A** representation there is an attraction of equal strength. This equality can be understood as follows. The diagrams of Fig. 1 serve dual purpose. In addition to showing contractions contributing to $E$, they can also represent the contributions to a direct calculation of pion scattering amplitudes. We refer to Fig. 1a as the gluon exchange amplitude, $T_g$, and to Fig. 1b as the quark exchange amplitude $T_q$. Following the standard usage for amplitudes, we *include* the Fermi-statistics sign in $T_q$, i.e. we use the opposite convention to that for $C$. Thus we find that $T(\mathbf{S}) = T_g + T_q$ and $T(\mathbf{A}) = T_g - T_q$. Now, it is possible to show that $T_g$ vanishes in the chiral limit [4], so that $T(\mathbf{S}) = T_q = -T(\mathbf{A})$.

Testing the relations Eq. (2.2) is particularly important for Wilson fermions, which we use here. This is because Wilson fermions explicitly break chiral symmetry, the violation only vanishing in the continuum limit. Thus we expect $T$ to contain a lattice artifact proportional to $\Lambda a/m_\pi^2$, where $a$ is the lattice spacing and $\Lambda$ is some non-perturbative scale. This artifact will dominate over the constant term in the chiral limit [7]. This is in contrast to staggered fermions where such artifacts are forbidden by the residual chiral symmetry [4]. With Wilson fermions one can isolate the physical result by doing the calculation at a number of values of the quark mass provided $f_\pi^2 T$ is a well behaved function of $m_\pi^2$. This remains to be checked. We find that, for $560\,\text{MeV} < m_\pi < 700\,MeV$ the chiral symmetry breaking effects are smaller than the statistical errors ($\sim 10\%$).

## 3. Calculational details

The energy of two pions in a finite box is obtained from the Euclidean correlator

$$C_{\pi\pi}(t) = \langle \sum_{\vec{x}_1} \mathcal{O}_1(\vec{x}_1,t) \sum_{\vec{x}_2} \mathcal{O}_2(\vec{x}_2,t)\, \mathcal{S}_3(\vec{x}_3,t\!=\!0)\, \mathcal{S}_4(\vec{x}_4,t\!=\!0) \rangle . \tag{3.1}$$

The sources $\mathcal{S}_i$ create the pions at $t = 0$, and the operators $\mathcal{O}_i$ (which we also call the "sinks") destroy them at time $t$. The representation of the two pion state is determined by the flavor of the sources and sinks. For example, we can select the $I = 2$ (or equivalently **S**) representation if both sources have the flavor of a $\pi^+$, and both $\mathcal{O}_1$ and $\mathcal{O}_2$ have the flavor of a $\pi^-$.

At large $|t|$ the correlator will fall as

$$C_{\pi\pi}(t) = Z_{\pi\pi}\exp(-E|t|) + \ldots , \tag{3.2}$$

where $E$ is the energy of the lightest two pion state. The ellipsis indicates contributions from excited states that are suppressed exponentially. This is similar to the behavior of the two point function used to calculate $m_\pi$

$$C_\pi(t) = \langle \sum_{\vec{x}_1} \mathcal{O}(\vec{x}_1,t)\, \mathcal{S}(\vec{x}_4,t\!=\!0) \rangle = Z_\pi \exp(-m_\pi|t|) + \ldots . \tag{3.3}$$

We take all quarks to be degenerate, so the flavor of the pion source $\mathcal{S}$ in this equation is unimportant; all that matters is that $\mathcal{O}$ has the conjugate flavor to $\mathcal{S}$. It is useful in practice to combine Eqs. (3.2) and (3.3)

$$\mathcal{R}(t) \equiv \frac{C_{\pi\pi}(t)}{C_\pi(t)^2} = \frac{Z_{\pi\pi}}{Z_\pi^2}\exp(-\delta E|t|) + \ldots , \tag{3.4}$$

and directly extract the energy shift $\delta E$.

The contractions which can contribute to $C_{\pi\pi}$ are shown in Fig. 1. The combination which is needed depends on the flavor of the two pion state. It is easy to see that only the Direct and Crossed contractions contribute for two pions in **S** or **A** representations [4]. We adopt the notation that $D(t)$ is a ratio as in Eq. (3.4), with the numerator being the Direct contraction. Similarly, $C(t)$ is the ratio with the Crossed contraction in the numerator. Then, as discussed above, and explained in more detail in Ref. [4], we can extract the energy shifts for the **S** and **A** representations using

$$D(t) + C(t) = Z_A \, e^{-\delta E_A |t|} , \qquad (3.5)$$

$$D(t) - C(t) = Z_S \, e^{-\delta E_S |t|} . \qquad (3.6)$$

The amplitudes $Z_S$ and $Z_A$ are shorthand for the ratios $Z_{\pi\pi}/Z_\pi^2$.

We are free to choose the form of the sources and sinks, as long as they both couple to a two pion state of the required flavor. This choice does not affect the value of the energy shift, but it does alter the signal to noise ratio. In order to improve this ratio, we should use sources with a large overlap with the lightest two pion state. In fact, as indicated in Eq. (3.1), we use the product of two independent single pion operators, and make no attempt to account for the correlations caused by the interactions between the pions. For the single pion sources (the $\mathcal{S}_i$ in Eq. (3.1)), we use wall and Wuppertal quark sources, which we have shown to be reasonably effective in producing single particle correlators [8] [9]. In addition, we use both pseudoscalar ($\mathcal{P} = \overline{\psi}\gamma_5\psi$) and axial vector ($\mathcal{A}_4 = \overline{\psi}\gamma_4\gamma_5\psi$) operators for each of the sources. There are thus four sources in all, which we label $W_P$ (wall with pseudoscalar), $W_A$ (wall with axial vector), $S_P$ (Wuppertal with pseudoscalar), and $S_A$ (Wuppertal with axial vector). For the sinks (the $\mathcal{O}_i$ in Eq. (3.1)) we use local operators, with Dirac structure either $P$ or $A$. Of the various possible combinations of sources and sinks we consider only those in which both sources are of the same type, and both sinks have the same Dirac structure as the sources. Thus we can label the ratios $\mathcal{R}(t)$ according to the choice of source, i.e. as $W_A$, $W_P$, etc. Finally, we always define the ratio $\mathcal{R}(t)$ with the same sources and operators in both the numerator and denominator. This is of no consequence for the energy shifts, but does affect the amplitudes $Z_{A,S}$, which we discuss further in Section 5.

We use 35 pure gauge configurations of size $16^3 \times 40$ generated at $\beta = 6.0$. We use Wilson fermions at two different quark masses, $\kappa = 0.154$ and $0.155$. The corresponding pions have masses of about 700 and 560 MeV respectively. The quark propagators are calculated on lattices doubled in time (i.e. of size $16^3 \times 80$), with periodic boundary conditions in all directions. We have previously used these lattices and propagators to study the spectrum and matrix elements [8][9], and we list the relevant results in Table 1. The value of $f_\pi$ given in Table 1 is different from that quoted in [8] for two reasons. First, we use the normalization such that the experimental value is $f_\pi = 93\ MeV$, and second we now use the mean field improved value 0.77 for the axial current renormalization constant [10], rather than the previous estimate 0.86. The statistical errors in individual data points shown in Figures 2-6 are calculated using the single elimination jack-knife procedure. The "forward" and "backward" propagators on the underlying $16^3 \times 40$ lattices give us two results for $R(t)$ on each lattice. Since these are correlated, we average them and treat them as a single result.

## 4. Results

To display our results, we use the quantity

$$\delta E_{\rm eff}(t) = \ln[\mathcal{R}(t)/\mathcal{R}(t+1)] \ . \tag{4.1}$$

This "effective energy shift" should reach a plateau of height $\delta E$ when $t$ is large enough that the lightest state dominates. Figures 2 and 3 show $\delta E_{\rm eff}(t)$ for the pseudoscalar operator $\mathcal{P}$ at $\kappa = 0.154$, using Wuppertal and Wall sources respectively. The results using the axial operator $\mathcal{A}_4$ are of poorer quality, and are not shown. There is a clear signal of a non-vanishing energy shift. We extract $\delta E$ by fitting $\mathcal{R}(t)$ to a single exponential, selecting the range of time-slices separately for each channel based on the extent of the plateau in the effective energy shift. The solid lines in the figures indicate the fit value over the range of the fit, while the dashed lines show the 1 $\sigma$ jack-knife errors.

It is apparent from Figs. 2 and 3 that the signal for Wall sources has smaller statistical errors than that for the Wuppertal sources. We can partly understand this as follows. The Wall source produces pions with $\vec{p} = 0$, while the Wuppertal source couples to pions having all possible momenta. Consequently the Wuppertal source correlators have an additional unwanted contribution at small $t$ from an excited state of two pions of equal and opposite momenta, each of magnitude $p = 2\pi/L$. For large volumes this state approaches the lightest state consisting of two pions both having $\vec{p} = 0$, and provides the largest contamination to $\mathcal{R}(t)$ rather than the states made up of radially excited pions. It may therefore become necessary to use wall sources for calculations on larger lattices.

In Figures 4-7 we show the ratios $\mathcal{R}(t)$ at $\kappa = 0.154$, for both $\mathcal{P}$ and $\mathcal{A}_4$ operators. These plots show that there is a contamination from "wrap-around" effects starting at $t \sim 30$. One of the two pions can propagate $N_t - t = 80 - t$ time-steps backwards, because of the periodic boundary conditions. This results in a contribution which is independent of $t$, but suppressed by roughly $\exp(-m_\pi N_t)/\exp(-2m_\pi t)$ compared to the forward propagation of the two pion state. In practice we always fit to time ranges satisfying $t_{\max} \leq 26$, for which we can ignore this contamination.

The results of our fits, together with the time ranges used, are given in Table 2. For both $S_P$ and $W_P$ correlators we fit using the full covariance matrix over the range of the plateau. We are unable to do this for the $S_A$ and $W_A$ channels, because some of the jackknife samples are too noisy. Our results for these channels are obtained keeping only the diagonal elements of the covariance matrix, i.e. neglecting correlations in $\mathcal{R}(t)$ between different time-slices. Because of this, we use only the $S_P$ and $W_P$ results to extract scattering amplitudes. The data show that the interaction in the $D + C$ channel is attractive, consistent with the result derived in the Appendix.

Our results for scattering amplitudes are presented in Table 3. We illustrate our procedure for obtaining these results using $T(S)$ as an example. For each jackknife sample we first average the fit value for $\delta E(S)$ from the $S_P$ and $W_P$ sources, and then solve the cubic polynomial given in Eq.(2.1) for $T(S)$. The central value and the error are given by the jackknife procedure, regarding the 35 data points as statistically independent. Within the same jackknife procedure we also calculate $T(A)$, and extract $T_q = (T(S) - T(A))/2$ and $T_g = (T(S) + T(A))/2$.

When solving Eq. (2.1) for $T(S)$ or $T(A)$ we monitor the effect of the $1/L^4$ and $1/L^5$ terms (using the values of $m_\pi$ given in Table 1). These turn out to be, respectively, $\sim 31\%$ and $8\%$ of the leading term. This suggests that the error introduced by truncating the series is a few percent in the scattering amplitudes. This is smaller than the statistical errors, which are approximately $10\%$.

To test the current algebra predictions, we calculate the combinations $4f_\pi^2 T$. These are included in Table 3, and shown in Fig. 8. The errors are calculated assuming that $f_\pi$ and $T$ are uncorrelated; we have checked on subsamples that the errors are similar if correlations are included. In the chiral limit the following relations should hold: $4f_\pi^2 T(S) = 1$, $T(S) = T_q = -T(A)$ and $T_g = 0$. Our results show that, within our errors, the second relation holds even at the relatively large quark masses we have used, and $4f_\pi^2 T$ lies between $0.76 - 0.88$. There is a small increase in $T$ between $\kappa = 0.154$ and $0.155$, but because of the size of the statistical errors we cannot conclude if this is related to the $1/m_\pi^2$ divergence expected for Wilson fermions in the chiral limit. Also, the errors are too large to extract a value for the gluon exchange amplitude $T_g$. All we can say is that it is much smaller than the quark exchange amplitude.

It is interesting to compare our results with those obtained using staggered fermions on the same lattices [4]. For technical reasons, we were able to calculate only the quantity $4Q = T_q - 2T_q T_g(c_1 m_\pi/4\pi L))$ with staggered fermions, and not $T_q$ and $T_g$ separately. If we use the values for $T_g$ obtained here, however, then we find that $4Q = T_q$ to good approximation. To compare the results, we note that the pion mass and $f_\pi$ match for the following parameter values: staggered $m_q = (0.02 + 0.03)$ with Wilson $\kappa = 0.154$, and staggered $m_q = (0.01 + 0.02)$ with $\kappa = 0.155$ [11]. The staggered results at these two masses are $T_q \approx 4Q = 67(8)$ and $78(6)$ respectively, which agree within errors with the results obtained here. It is reassuring to find that the two formulations, each with their separate technical problems, yield mutually consistent results.

## 5. Expected behavior of $\mathcal{R}(t)$ with Wilson Fermions

As shown in Table 2 and Figs. 4-7, different source and sink operators give considerably different values for $Z_{A,S}$. This variation is more marked for smaller quark mass. It turns out that the values for $Z_{A,S}$ can be understood semi-quantitatively using a simple model, as we explain in this section. This model is similar to that used in the analysis of our staggered fermion data [4].

We begin by imagining that the source creates two pions each having $\vec{p} = 0$. The pion operators could be local or smeared, but should have a finite extent that is much smaller than the lattice size. Also, the same source operators are used in the numerator and denominator of Eq. (3.4). In this case we expect that $Z_{A,S} \to 1$ as $L \to \infty$, because the lightest two pion state differs from two independent, zero momentum pions by terms which vanish as $L \to \infty$. Thus the $Z$-factors in the numerator and denominator of Eq. (3.4) cancel. Assuming $Z_{A,S} = 1$, and using Eqs. (3.5) and (3.6), the behavior of the direct and crossed contractions is

$$\begin{aligned} D(t) &= 1 - \frac{E_S + E_A}{2} t + O(t^2)\,, \\ C(t) &= -\frac{E_S - E_A}{2} t + O(t^2)\,, \end{aligned} \quad (5.1)$$

for $t$ large enough that the contributions of excited states have died away. What is important here is that only $D(t)$ is non-zero when extrapolated back to $t = 0$; the constant term corresponds to the two pions propagating from source to sink without interactions, which is only possible in the Direct channel. In the Crossed channel, at least one quark exchange interaction is required. This gives rise to the term linear in $t$, since the interaction can occur at any time. The term linear in $t$ in $D(t)$ is due to the gluon exchange interaction.

In practice $Z_A$ and $Z_S$ differ from unity, for two reasons. First, the two pion state is altered by interactions. This gives corrections proportional to $1/L^2$ [4], which we assume are small and ignore. Second, our sources are not two independent pion operators each having $\vec{p} = 0$, but rather a single wall or Wuppertal quark source. This gives large corrections to $Z_{A,S}$, due to the overlap, which do not vanish as $L \to \infty$, and it is these which we estimate.

In our set up a state of two quarks and two antiquarks, all in close proximity to one another, is created rather than two separate pions. We attempt to pair the quarks and antiquarks into pions by making two color singlets, each with pseudoscalar quantum numbers. But, by performing a combined color and Dirac Fierz transformation, we find that we are also creating, with non-vanishing amplitude, two pions with the opposite $\overline{q}q$ pairings. Explicitly, the Fierz transformations are

$$\begin{aligned}
\mathcal{P} \otimes \mathcal{P} &\to \frac{1}{12}\left(\mathcal{P} \otimes \mathcal{P} + \mathcal{A}_4 \otimes \mathcal{A}_4 \ldots \right), \\
\mathcal{A}_4 \otimes \mathcal{A}_4 &\to \frac{1}{12}\left(\mathcal{P} \otimes \mathcal{P} + \mathcal{A}_4 \otimes \mathcal{A}_4 \ldots \right),
\end{aligned} \tag{5.2}$$

where $\mathcal{P} \otimes \mathcal{P} \equiv (\overline{\psi}_1 \gamma_5 \psi_2)(\overline{\psi}_3 \gamma_5 \psi_4)$ and $\mathcal{A}_4 \otimes \mathcal{A}_4 \equiv (\overline{\psi}_1 \gamma_4 \gamma_5 \psi_2)(\overline{\psi}_3 \gamma_4 \gamma_5 \psi_4)$, parentheses implying spin and color traces. The sign due to fermion exchange is not included in these Fierz identities, since we do not include the sign in our definitions of $D(t)$ and $C(t)$. The Fierz relations hold for both wall and Wuppertal sources, in the latter case because all the products of link matrices begin at the same site. We have shown only the $\mathcal{P} \otimes \mathcal{P}$ and $\mathcal{A}_4 \otimes \mathcal{A}_4$ parts of the Fierzed combinations because the correlators of these operators have the dominant contribution. Other tensor structures give no contribution for two separated pion sources in infinite volume, and thus give contributions here that are suppressed by powers of $1/L$. The same is true for operators consisting of two color octets. The magnitude of these neglected terms can be significant, especially for Wuppertal sources, as shown by the difference between our data and the estimates presented below.

There is no large Fierz contribution from the sinks, since these do consist of two independent pion operators, each having $\vec{p} = 0$. The only contribution occurs when the two operators overlap, and is suppressed by powers of $1/L$.

The Fierz relations mean that our Crossed contractions contain a part in which the quarks have already been exchanged before we can identify the state as one with two pions, so that no subsequent quark-exchange interaction is necessary. This leads to a constant term in the Crossed contraction. The Fierz contributions to the Direct contraction do not, however, affect the constant term, for there must be an additional quark exchange interaction to bring the quarks and antiquarks back to their original pairings. This discussion motivates the following assumptions for the constant terms

$$D(t=0)_{P,A} \approx 1 ,$$
$$C(t=0)_P \approx \frac{1}{12}(1 + C_A^2/C_P^2) , \qquad (5.3)$$
$$C(t=0)_A \approx \frac{1}{12}(C_P^2/C_A^2 + 1) .$$

The subscript indicates the type of correlator, and the constants $C_A$, $C_P$ are the amplitudes for creating single pions with $\vec{p} = 0$ using the operators $\mathcal{A}_4$ and $\mathcal{P}$ respectively. The ratio $C_P/C_A$ is 2.5 and 3.1 for Wuppertal sources, and 1.6 and 1.9 for wall sources, at $\kappa = 0.154$ and 0.155, respectively. Using these values, we can calculate the $Z$'s using

$$Z_S \approx D(t=0) - C(t=0) , \qquad Z_A \approx D(t=0) + C(t=0) . \qquad (5.4)$$

The predictions are collected in Table 4. They give a good semi-quantitative representation of the data for $Z_{A,S}$ in Table 2. In particular, we can understand the small value of $Z_S$ for the $S_A$ operators as being due to a large cancellation between the Direct contraction and the Fierz contribution to the Crossed contraction, the latter being enhanced by the large ratio $C_P/C_A$ for Wuppertal sources. This cancellation is most likely why the signal is so noisy in this channel.

## 6. Conclusions

We find that it is straightforward to calculate the finite volume energy shift for channels not involving $\overline{q}q$ annihilation. The calculation is much less involved using Wilson fermions than that we carried out with staggered fermions [4]. We are able to work on a lattice of modest size ($L \approx 1.6\text{fm}$) because the interactions in the channels we consider are relatively weak. From the energy shifts we extract the quark exchange amplitude, and place a bound on the gluon exchange amplitude. Our results are consistent with the predictions of current algebra, on the other hand the quarks used in the calculation are not light enough to expose the expected artifacts due to the breaking of chiral symmetry by Wilson fermions.

It is important to extend this work to smaller quark masses, where the divergence in $T$ due to chiral symmetry breaking should show up. Furthermore, the result should be checked on a larger volume to verify that the asymptotic form of the finite volume dependence can be used.

## Acknowledgments

The $16^3 \times 40$ lattices were generated at NERSC at Livermore using a DOE allocation. The calculation of quark propagators and the analysis has been done at the Pittsburgh Supercomputing Center, San Diego Supercomputer Center, NERSC and Los Alamos National Laboratory. We are very grateful to Jeff Mandula, Norm Morse, Ralph Roskies, Charlie Slocomb and Andy White for their support of this project. AP thanks Los Alamos National Laboratory for hospitality during the course of this work. SRS is supported in part by DOE contract DE-AC05-84ER40150 and by an Alfred P. Sloan Fellowship.

### Appendix A. Correlation Inequalities for Scattering Amplitudes

One can derive rigorous inequalities among correlation functions for vector-like gauge theories such as QCD. The basis of such inequalities is the positivity of the measure in the Euclidean path integral:

$$[dA_\mu][d\overline{\psi}][d\psi] \exp[-S_{gauge} - S_{fermion}] \geq 0 \ . \tag{A.1}$$

This property has been exploited, both on the lattice [12] and in the continuum [13], to derive inequalities among 2−point correlation functions. As a result it was shown that the pion is the lightest meson. Here we apply the same arguments to 4−point correlation functions in finite volume to derive constraints on $\pi - \pi$ scattering amplitudes. The only other work extending the derivation of inequalities to higher order correlation functions that we are aware of is Ref. [14], where it is shown that the pion wavefunction is largest at $\vec{r} = 0$ using a 4−point correlation function.

We consider the 4−point correlation functions corresponding to the $\pi - \pi$ scattering at threshold. Let the sources for the two pion be at time $t = 0$ and the two pion sinks be at time $t = T$. We take all four pion operators to be point-like with Dirac structure $\overline{\psi}\gamma_5\psi$ and zero 3−momentum. In terms of the quark propagator $G(\vec{x}, 0; \vec{y}, T)$, the Direct and Crossed correlators are:

$$\begin{aligned}
D(0, T; \vec{p} = 0) &= \sum_{\vec{x},\vec{y},\vec{z},\vec{w}} \langle \text{Tr}(G(\vec{x}, 0; \vec{y}, T)G^\dagger(\vec{x}, 0; \vec{y}, T)) \text{Tr}(G(\vec{z}, 0; \vec{w}, T)G^\dagger(\vec{z}, 0; \vec{w}, T)) \rangle \ , \\
C(0, T; \vec{p} = 0) &= \sum_{\vec{x},\vec{y},\vec{z},\vec{w}} \langle \text{Tr}(G(\vec{x}, 0; \vec{y}, T)G^\dagger(\vec{z}, 0; \vec{y}, T)G(\vec{z}, 0; \vec{w}, T)G^\dagger(\vec{x}, 0; \vec{w}, T)) \rangle \ ,
\end{aligned}$$
(A.2)

where the trace is taken over the color and spin indices.

Using the Schwarz inequality

$$\langle f f^\dagger \rangle \geq |\langle f \rangle|^2 \ , \tag{A.3}$$

we get the relation

$$D(0, T; \vec{p} = 0) \geq [P(0, T; \vec{p} = 0)]^2 \tag{A.4}$$

where the zero 3−momentum pion correlator,

$$P(0, T; \vec{p} = 0) = \sum_{\vec{x},\vec{y}} \langle \text{Tr}(G(\vec{x}, 0; \vec{y}, T)G^\dagger(\vec{x}, 0; \vec{y}, T)) \rangle \ , \tag{A.5}$$

is by itself positive definite. This inequality implies that the contribution of this diagram to the two pion interaction is attractive.

These results can be generalized to other Dirac structures. In fact the interaction between any two mesons, e.g. two rhos, is attractive in the Direct channel. Note that the derivation of this inequality did not depend on the volume of the system. This is analogous to the zero temperature mass inequalities of Ref. [12][13] which can be used at finite temperature to give relations between hadronic screening lengths.

The crossed correlator given in Eq. (A.2) can be rewritten as

$$C(0, T; \vec{p} = 0) = \sum_{\vec{x},\vec{z}} \langle \text{Tr}([\sum_{\vec{y}} G(\vec{x}, 0; \vec{y}, T)G^\dagger(\vec{z}, 0; \vec{y}, T)][\sum_{\vec{w}} G(\vec{x}, 0; \vec{w}, T)G^\dagger(\vec{z}, 0; \vec{w}, T)]^\dagger) \rangle \ , \tag{A.6}$$

and is therefore also positive. Combining this fact with the inequality in Eq. (A.4) shows that the scattering amplitude in the flavor antisymmetric channel, corresponding to the combination $D + C$, is

$$D(0, T; \vec{p} = 0) + C(0, T; \vec{p} = 0) \geq [P(0, T; \vec{p} = 0)]^2 \ . \tag{A.7}$$

This inequality on the correlators implies that the exponential fall-off with time in this channel is slower than that for two non-interacting pions. It follows that the interaction energy $\delta E$ in the antisymmetric channel is negative, i.e. the scattering length is positive.

## References

[1] M. Lüscher, *Commun. Math. Phys.* **104** (1986) 177; **105** (1986) 153

[2] M. Lüscher, *Nucl. Phys.* **B354** (1991) 531

[3] H.W. Hamber, E. Marinari, G. Parisi and C. Rebbi, *Nucl. Phys.* **B225** (1983) 475

[4] S. Sharpe, R. Gupta and G. Kilcup, *Nucl. Phys.* **B383** (1992) 309

[5] S. Weinberg, *Phys. Rev. Lett.* **17** (1966) 616

[6] M. Guagnelli, E. Marinari and G. Parisi, *Phys. Lett.* **240B** (1990) 188

[7] N. Kawamoto and J. Smit, *Nucl. Phys.* **B192** (1981) 100 ;
J. Hoek, N. Kawamoto, J. Smit, *Nucl. Phys.* **B199** (1982) 495

[8] D. Daniel, R. Gupta, G. Kilcup, A. Patel, S. Sharpe, *Phys. Rev.* **D46** (1992) 46

[9] R. Gupta, D. Daniel, G. Kilcup, A. Patel, S. Sharpe, LAUR-91-3522

[10] G.P. Lepage and P. Mackenzie, Int. Symp. *"LATTICE 90"*, Proceedings of the International Conference on Lattice Field Theory, Tallahassee, Florida, 1990, Eds. U. M. Heller *et al.*, *Nucl. Phys.* **B** (*Proc. Suppl.*) **20**, (1991) 173; ;
preprint FERMILAB-PUB-19/355-T (9/92)

[11] R. Gupta, G. Guralnik, G. Kilcup and S. Sharpe, *Phys. Rev.* **D43** (1991) 2003.

[12] D. Weingarten, *Phys. Rev. Lett.* **51** (1983) 1830.

[13] C. Vafa and E. Witten, *Nucl. Phys.* **B234** (1984) 173;
E. Witten, *Phys. Rev. Lett.* **51** (1983) 2351.

[14] S. Nussinov and M. Spiegelglas, *Phys. Lett.* **202B** (1988) 265

| $\beta$ | $\kappa$ | $Lattice$ | $N_{conf}$ | $m_\pi$ | $f_\pi$ |
|---|---|---|---|---|---|
| 6.0 | 0.154 | $16^3 \times 80$ | 35 | 0.364(6) | 0.057(3) |
| 6.0 | 0.155 | $16^3 \times 80$ | 35 | 0.297(9) | 0.055(3) |

**Table 1.** Summary of results from Ref. [8] needed in this calculation. $f_\pi$ is normalized such that the experimental value is $f_\pi = 93 \ MeV$, and is obtained using the mean field improved value 0.77 for the axial current renormalization constant.

| | $D + C$ | | | $D - C$ | | |
|---|---|---|---|---|---|---|
| Correlator | $Fit$ | $Z_A$ | $-E_A$ | $Fit$ | $Z_S$ | $E_S$ |
| $\kappa = 0.154$ | | | | | | |
| $S_P$ | $9 - 16$ | 1.19(4) | 0.020(2) | $8 - 25$ | 0.93(7) | 0.022(6) |
| $W_P$ | $8 - 16$ | 1.10(2) | 0.022(3) | $8 - 15$ | 0.92(2) | 0.018(4) |
| $S_A$ | $8 - 16$ | 1.77(10) | 0.019(7) | $8 - 20$ | 0.46(7) | 0.019(10) |
| $W_A$ | $8 - 15$ | 1.30(5) | 0.025(6) | $8 - 26$ | 0.78(7) | 0.016(9) |
| $\kappa = 0.155$ | | | | | | |
| $S_P$ | $8 - 17$ | 1.23(6) | 0.022(6) | $8 - 18$ | 0.87(6) | 0.022(8) |
| $W_P$ | $8 - 18$ | 1.14(3) | 0.027(3) | $8 - 15$ | 0.92(3) | 0.024(5) |
| $S_A$ | $8 - 15$ | 2.07(17) | 0.020(9) | $8 - 20$ | 0.27(11) | 0.023(25) |
| $W_A$ | $8 - 15$ | 1.49(9) | 0.023(5) | $7 - 20$ | 0.67(8) | 0.017(13) |

**Table 2.** Results for the amplitude and energy shifts obtained from fits to $\mathcal{R}(t)$ of correlators for the $D \pm C$ channels. The four kinds of correlators, $S_P$, $W_P$, $S_A$ and $W_A$ are described in the text. We also give the range of time-slices over which the fit is made.

|  | $\kappa = 0.154$ | $\kappa = 0.155$ |
|---|---|---|
| $T(S)$ | 58.7(8.7) | 69.6(10.7) |
| $T(A)$ | −61.8(6.5) | −73.0( 6.6) |
| $T_q$ | 60.2(4.6) | 71.3( 6.6) |
| $T_g$ | −1.6(5.0) | −1.7( 6.0) |
| $4f_\pi^2 T(S)$ | 0.76(14) | 0.84(16) |
| $4f_\pi^2 T(A)$ | −0.80(12) | −0.88(13) |
| $4f_\pi^2 T_q$ | 0.78(10) | 0.86(12) |
| $4f_\pi^2 T_g$ | −0.02( 7) | −0.02( 7) |

**Table 3.** Final results for the scattering amplitudes.

| Correlator | $Z_A$ | $Z_S$ |
|---|---|---|
| $\kappa = 0.154$ | | |
| $S_P$ | 1.10 | 0.90 |
| $W_P$ | 1.12 | 0.88 |
| $S_A$ | 1.60 | 0.40 |
| $W_A$ | 1.30 | 0.70 |
| $\kappa = 0.155$ | | |
| $S_P$ | 1.09 | 0.91 |
| $W_P$ | 1.11 | 0.89 |
| $S_A$ | 1.88 | 0.12 |
| $W_A$ | 1.38 | 0.62 |

**Table 4.** Model predictions for the intercepts of $\mathcal{R}(t)$.

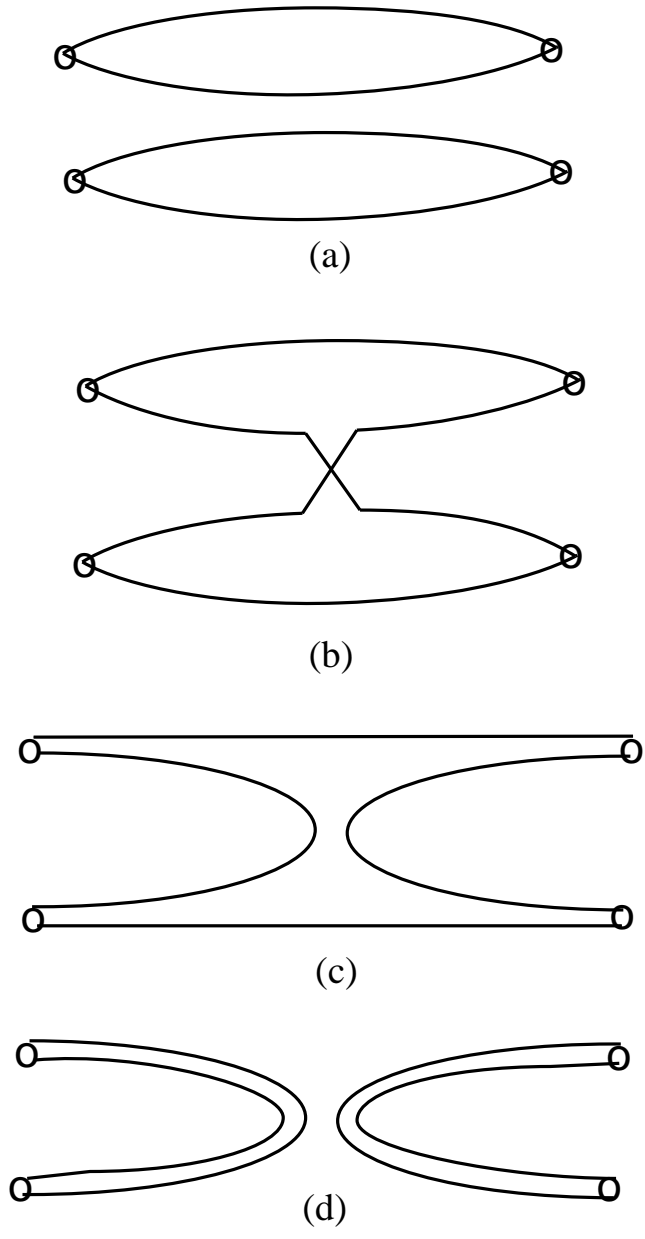

**Fig. 1.** The four different contractions that contribute to the two pion correlator: a) Direct, or gluon exchange, channel ($D$), b) Crossed, or quark exchange, channel ($C$), c) Single Annihilation and d) Double Annihilation. The diagrams also correspond to the amplitudes that contribute to $\pi\pi$ scattering.

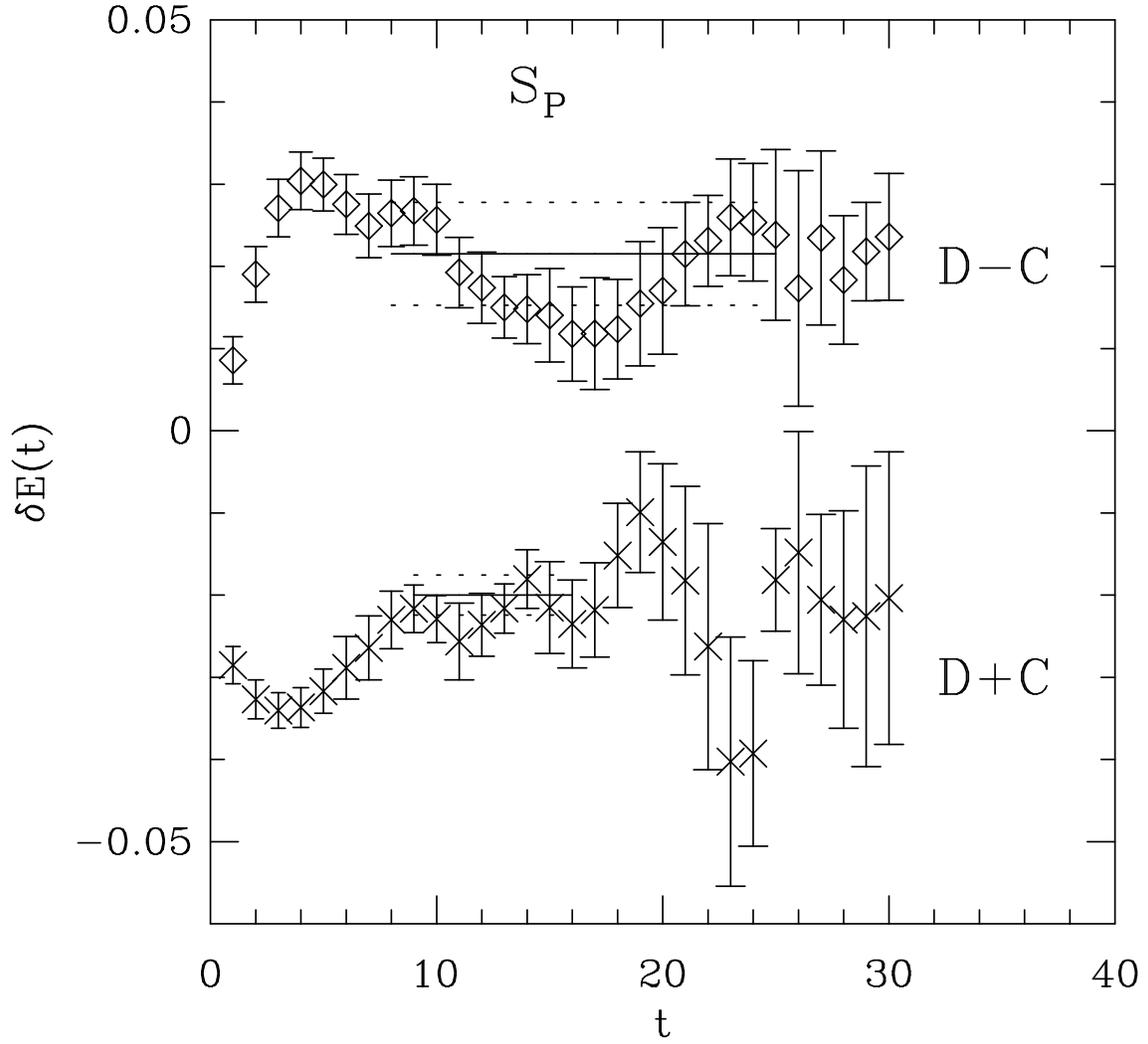

**Fig. 2** The effective energy shift $\delta E(t)$ using $S_P$ correlators at $\kappa = 0.154$, for both the $I = 2$ (**S**) representation $(D - C)$, and the **A** representation $(D + C)$.

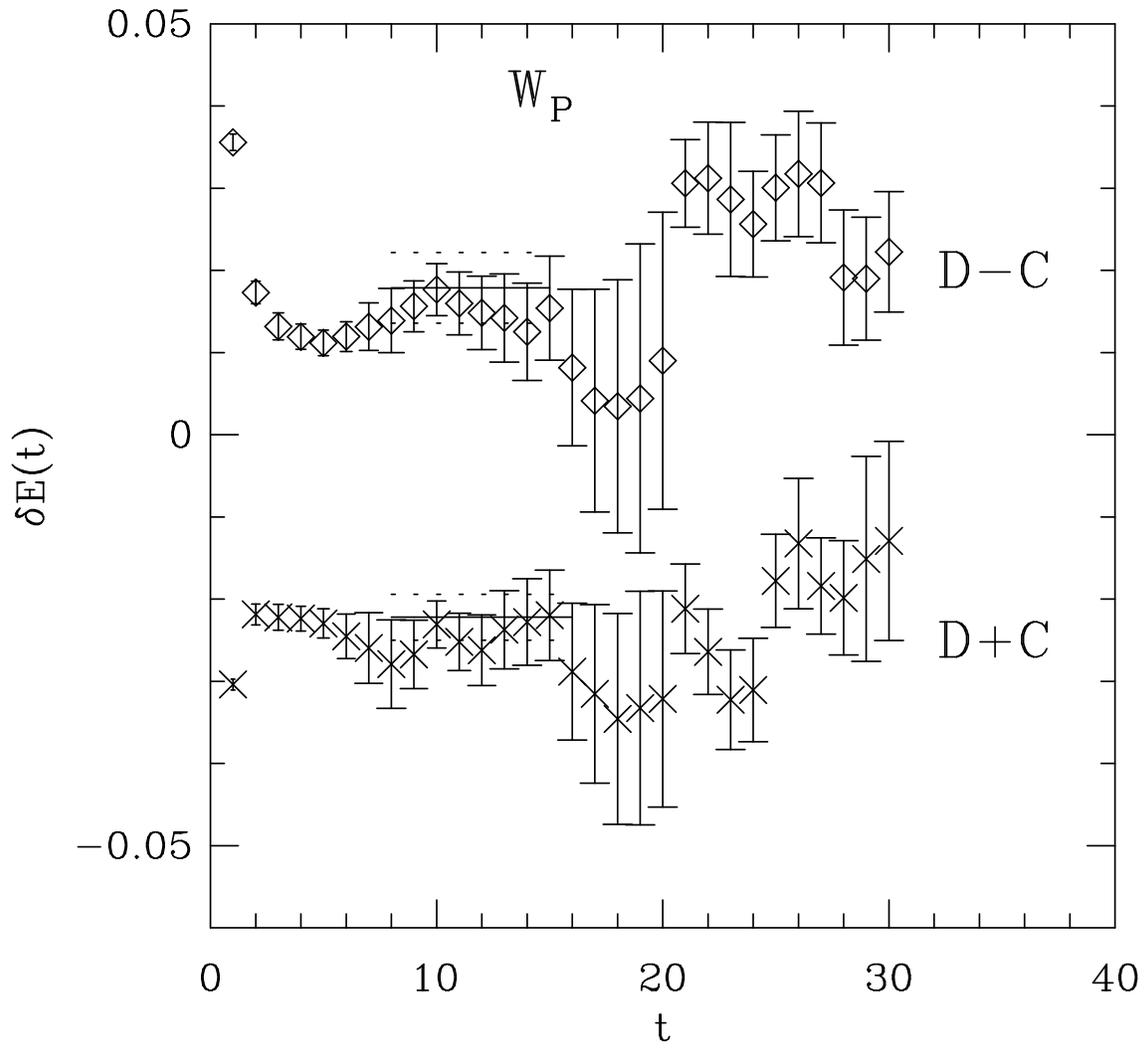

**Fig. 3** As in Fig. 2 but for the $W_P$ correlators.

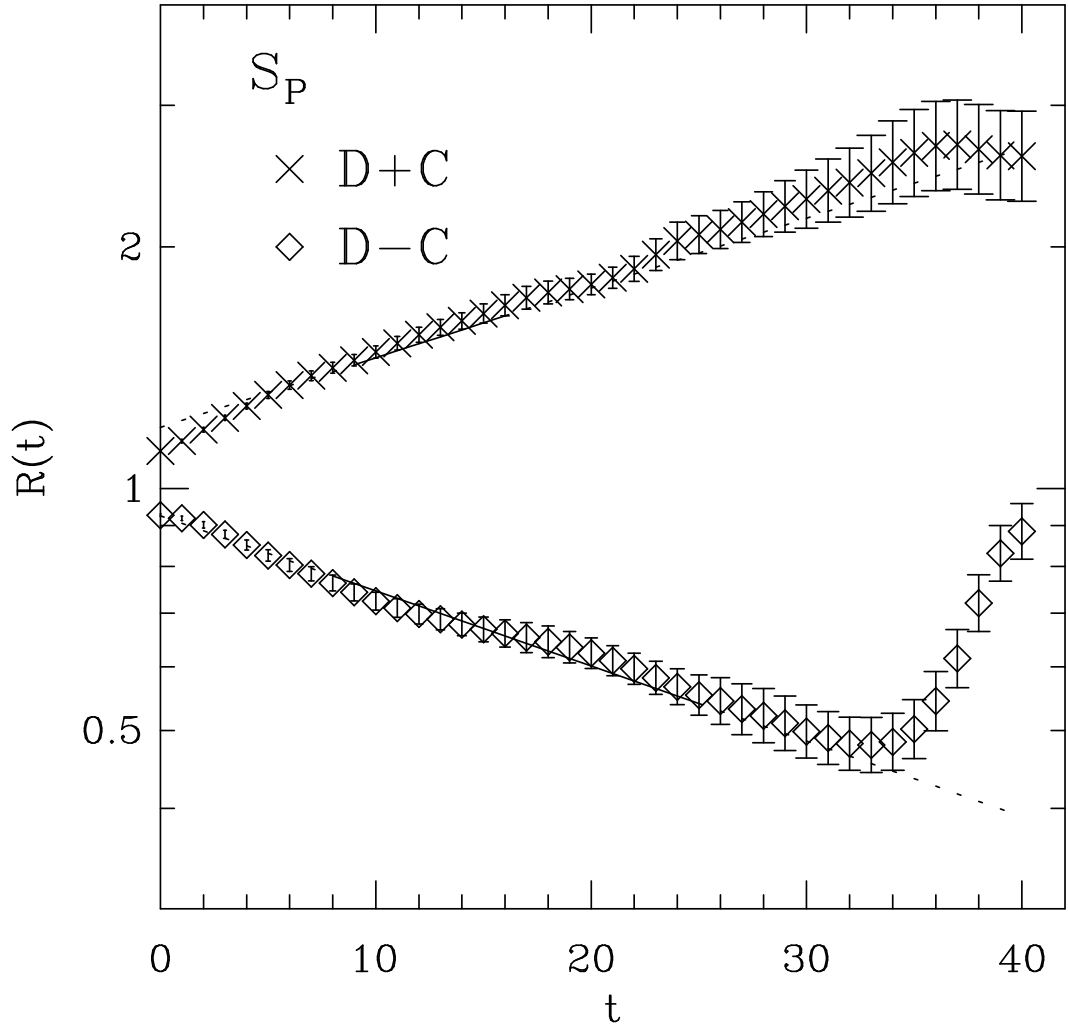

**Fig. 4.** $\mathcal{R}(t)$ using $S_P$ correlators at $\kappa = 0.154$.

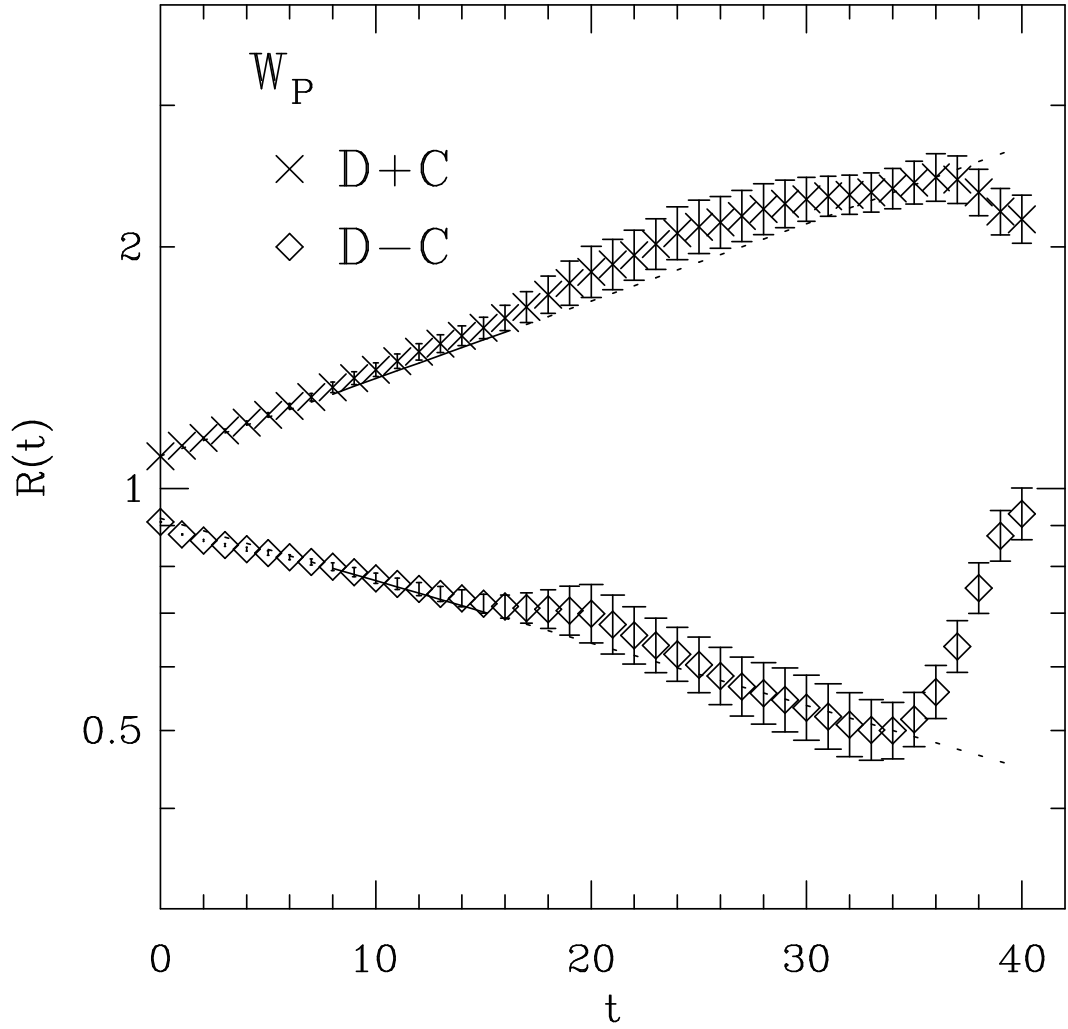

**Fig. 5.** $\mathcal{R}(t)$ using $W_P$ correlators at $\kappa = 0.154$.

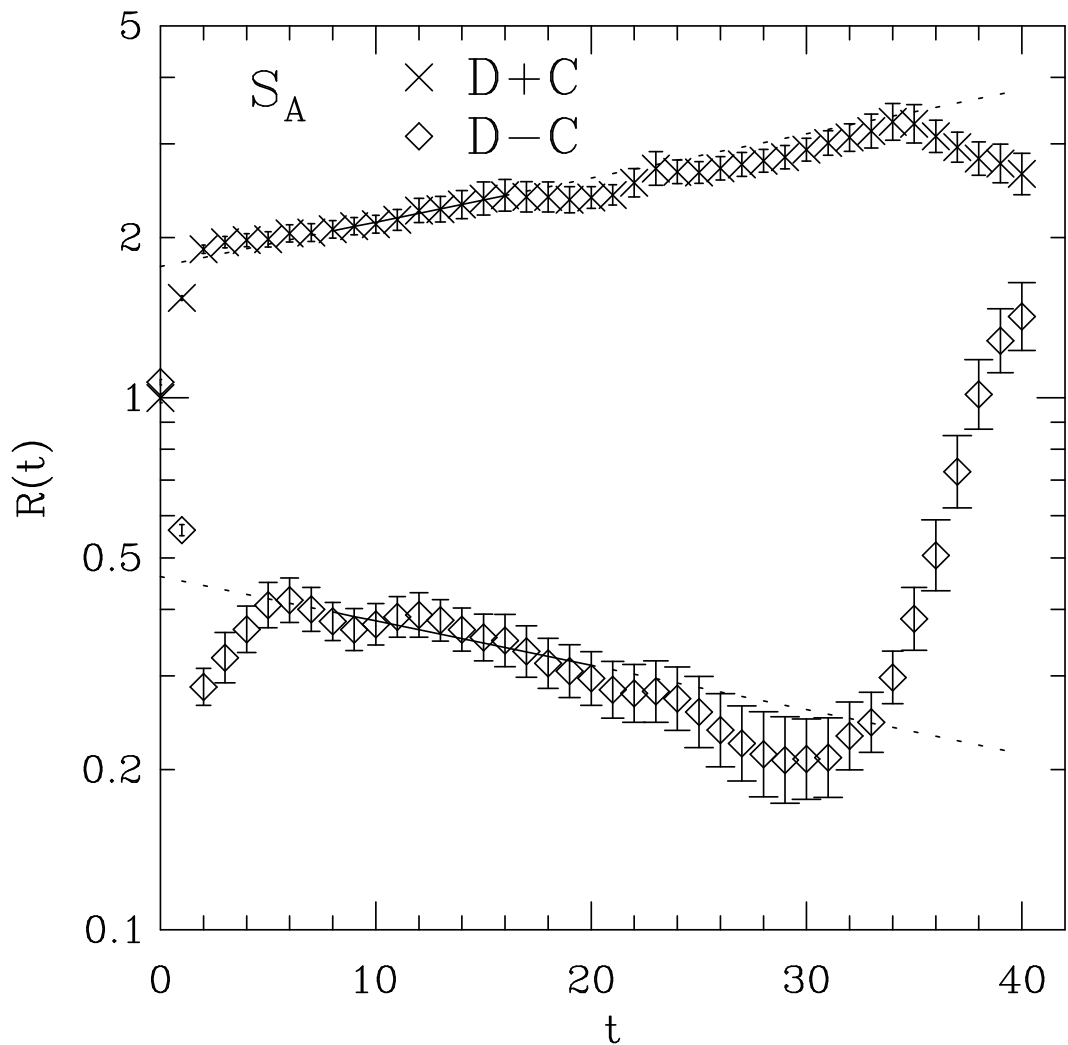

**Fig. 6.** $\mathcal{R}(t)$ using $S_A$ correlators at $\kappa = 0.154$.

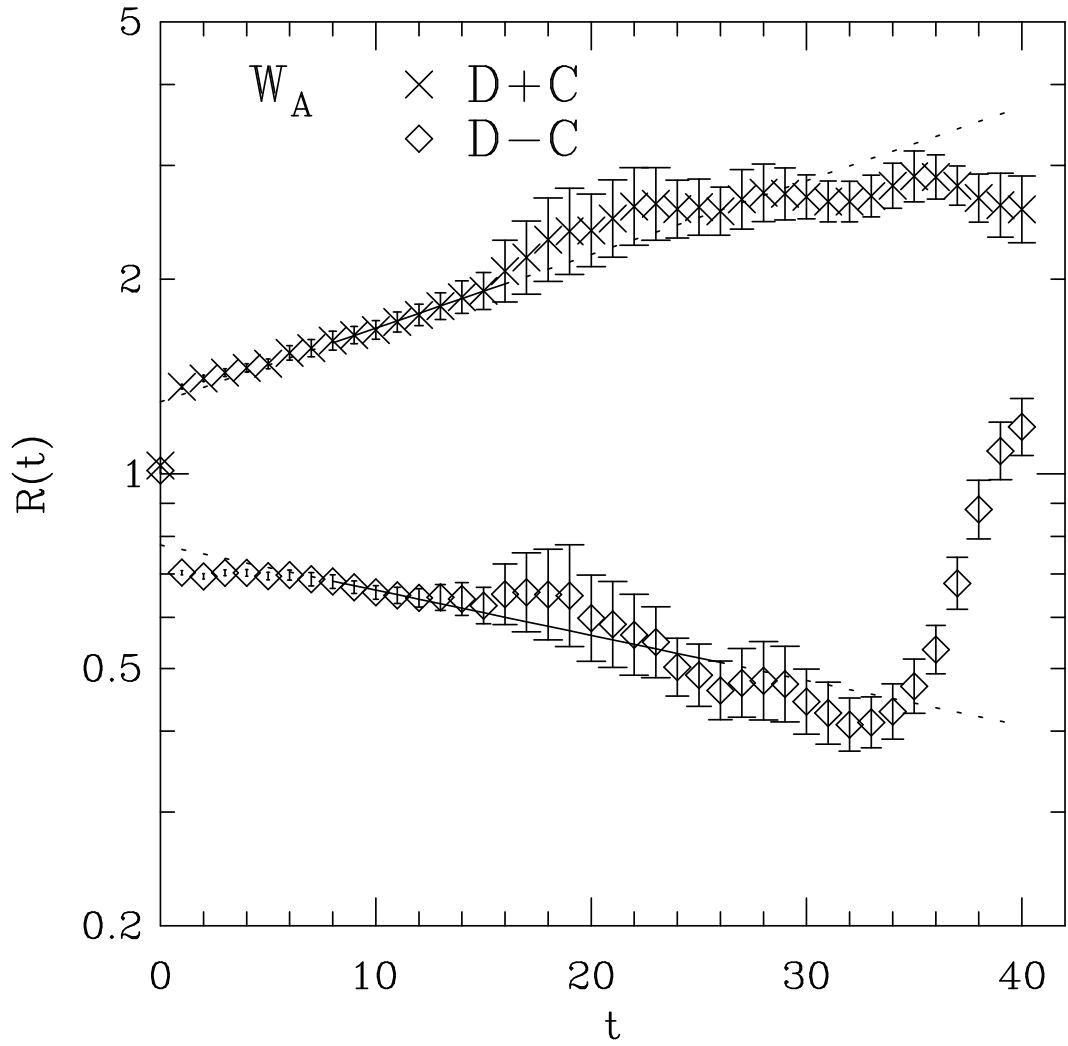

**Fig. 7.** $\mathcal{R}(t)$ using $W_A$ correlators at $\kappa = 0.154$.

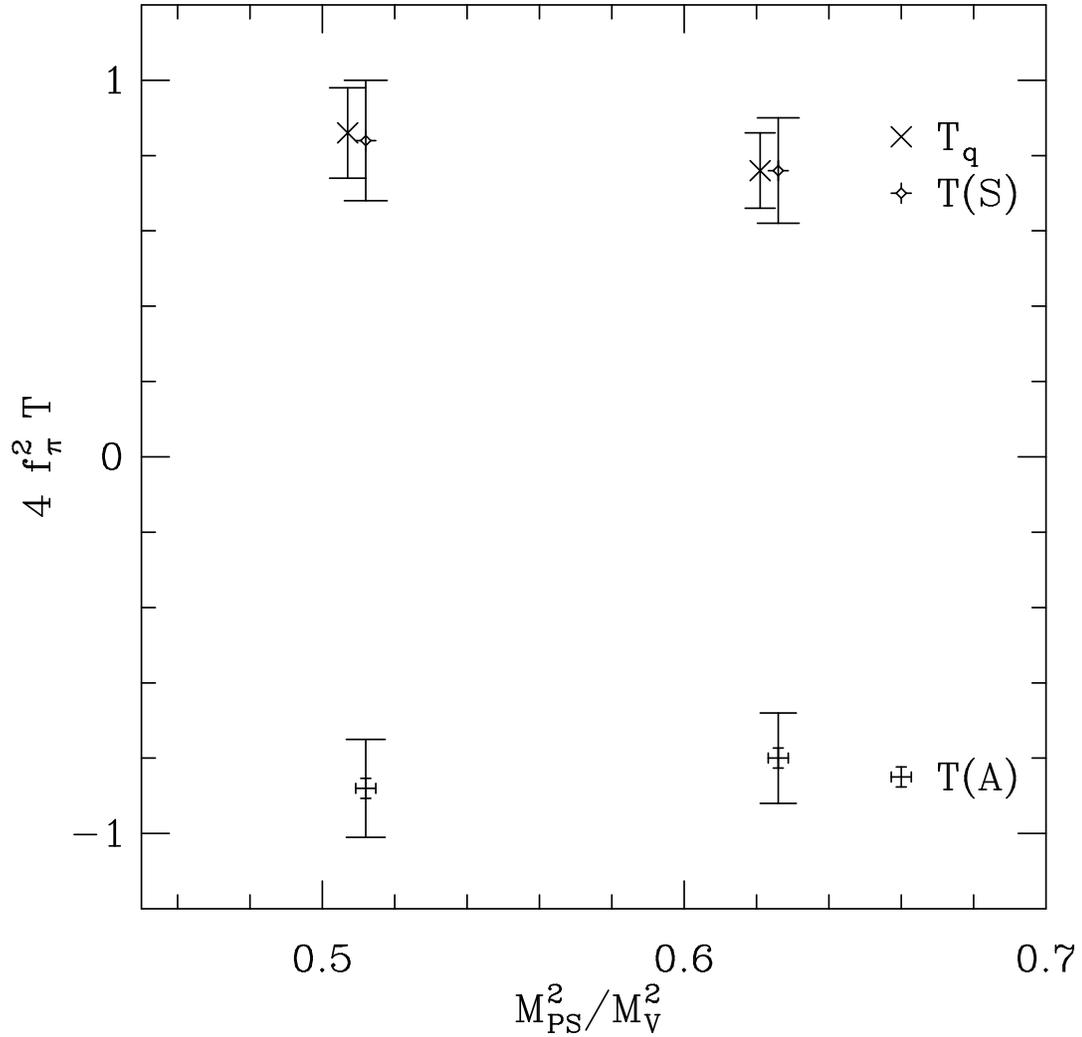

**Fig. 8.** $4f_\pi^2 T$ plotted versus $m_\pi^2/m_\rho^2$ to test the chiral behavior. The data for $T_q$ has been displaced by $-0.05$ along the x-axis for clarity.